# Sum-Frequency Generation Spectro-Microscopy in the Reststrahlen Band of Wurtzite-type Aluminum Nitride

D. S. Mader[1], R. Niemann[1], M. Wolf[1], S. F. Maehrlein[1], and A. Paarmann[1,*]

[1]*Fritz Haber Institute of the Max Planck Society, Faradayweg 4-6, 14195 Berlin, Germany*

*Corresponding author: alexander.paarmann@fhi-berlin.mpg.de

**Abstract** Nonlinear-optical microscopy and spectroscopy provide detailed spatial and spectroscopic contrast, specifically sensitive to structural symmetry and order. Ferroics, in particular, have been widely studied using second harmonic generation imaging, which provides detailed information on domain structures but typically lacks spectroscopic detail. In contrast, infrared-visible sum-frequency generation (SFG) spectroscopy reveals details of the atomic structure and bonding via vibrational resonances, but conventionally lacks spatial information. In this work, we combine the benefits of nonlinear optical imaging and SFG spectroscopy by employing SFG spectro-microscopy using an infrared free-electron laser. Specifically, we demonstrate the feasibility of SFG spectro-microscopy for spectroscopy using in-plane anisotropic wurtzite-type aluminum nitride as a model system. We find the experimental spectra to agree well with our theoretical calculations and we show the potential of our microscope to provide spatially resolved spectroscopic information in inhomogeneous systems such as ferroics and their domains in the near future.

## INTRODUCTION

Non-centrosymmetric materials especially ferroics show a broad range of technological applications from the nano to the wafer scale. For example, ferroelectric thin films can be integrated on nanoscale devices on Si chips, while ferroelectric single crystals are used for amplitude modulators, phase modulators or electro-optic shutters[1,2]. Properties of ferroelectric domains, but also domain walls, are employed for memory storage devices[3,4]. Therefore, due the potential for optical memory devices[5] the optical control of ferroelectric and/or magnetic polarizations remains an ongoing field of research[6,7]. Lastly, the characterization of new multiferroic materials reveals new material properties and opportunities[8].

A powerful tool to study such non-centrosymmetric systems is second harmonic generation (SHG) microscopy[9,10]. As an even-order nonlinear effect, SHG directly reflects on the broken inversion symmetry of a system. In ferroic systems, the sensitivity of SHG to the incident and detected polarization often provides an intrinsic domain contrast. In most cases, however, SHG imaging is employed non-resonantly, thus lacking access to the specifics of the local lattice structure beyond the symmetry properties.

In contrast, infrared-visible (IR-VIS) sum-frequency generation (SFG) spectroscopy is an alternative second-order nonlinear optical technique, commonly employed resonantly to reveal vibrational details of non-centrosymmetric media and interfaces. Nowadays, SFG is predominantly used in molecular, biological and solid-state research to study surfaces and interfaces[11-13]. However, in 1966 the first reported SFG work by W. L. Faust and C. H. Henry[14] investigated phonon resonances in bulk non-centrosymmetric gallium phosphide. In a further prominent example, W.-T. Liu and Y. R. Shen showed the underlying mathematical theory behind SFG spectroscopy and applied it to bulk $\alpha$-quartz[15].

In IR-VIS SFG spectroscopy, the material is excited near-resonance in the IR. The IR-induced polarization is mixed with a non-resonant VIS beam leading to generation of visible sum-frequency light. Compared to SHG, IR-VIS SFG benefits from the resonant enhancement in the IR that is also specific to the local order and bonding of the material. In contrast to IR spectroscopy, SFG benefits from the sensitivity and resolution of detectors and cameras in the VIS compared to the IR. Even though SFG spectroscopy is commonly used in solid-state physics, e.g., to probe phonon resonances, it has been rarely used for imaging.

Here, we merge nonlinear optical imaging with SFG spectroscopy to obtain SFG spectro-microscopy[16,17]. We implement IR-VIS SFG spectro-microscopy by using intense IR radiation from a free-electron laser (FEL). The first proof-of-concept by Niemann et al. applied this method to nanopillar patches on in-plane isotropic SiC to focus on the spatial resolution of the microscope[18]. Here, we complementarily extend this proof-of-concept to the spectroscopic capabilities of the SFG microscope using a prototype in-plane anisotropic bulk material. Specifically, we measure the spectral evolution of *m*-plane wurtzite-type aluminum nitride (AlN) for 8 different combinations of IR and VIS polarizations and sample orientations. The suitability of our method for simultaneous spectroscopy and imaging of non-centrosymmetric materials is proven by the good match between our experiments and theoretical calculations as well as the microscopically resolved features. In the near future our method will be applied to ferroics and their domains.

## EXPERIMENT

The geometry of the SFG microscope in its lab coordinates is shown in Figure **1a**. The sample surface is oriented in the *x-y*

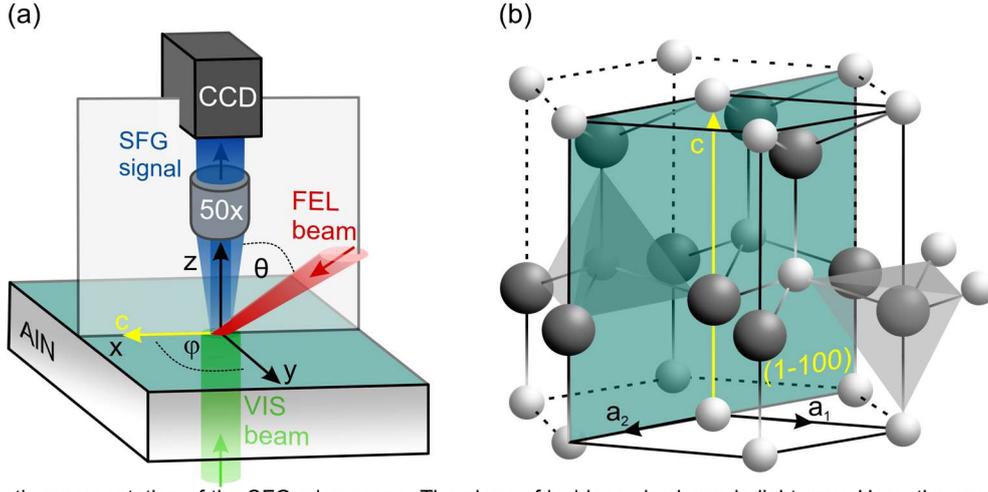

**Figure 1:** (a) Schematic representation of the SFG microscope. The plane of incidence is shown in light grey. Here, the crystal axis *c* is parallel to *x*. (b) Unit cell of wurzite-type aluminum nitride. The surface orientation for an *m*-plane (1-100) crystal is marked in green. The sample surface is oriented perpendicular to the plane of incidence. The optical axis *c* is marked in yellow.

plane with its normal along the *z*-axis. The IR light is incident on the frontside in the *x-z* plane, where its electric field vector oscillates in the *x-z* plane for parallel (p) and along *y* for perpendicular (s) polarization. The angle of incidence is kept at $\theta = 50°$ throughout the study. The VIS laser illuminates the sample from the backside under normal incidence such that the VIS electric field is oriented along *x* (labelled as p) and along *y* (labelled as s). The SFG light propagates along *z* towards the microscope objective without selection of a specific SFG polarization component.

We use the FEL at the Fritz Haber Institute (FHI-FEL) as a tunable IR light source in the frequency range between 600 and 1050 cm$^{-1}$, with a narrow bandwidth below 0.7%. The FEL generates macro pulses with an energy of ~5 mJ and a duration of ~8 μs at a repetition rate of 10 Hz. Each macro pulse contains micro pulses at 55 MHz repetition rate. Due to the non-normal incidence, the IR-irradiated area on the sample is elliptical, roughly 500 μm by 360 μm, leading to a uniform illumination of the camera's 275 x 275 μm² field of view. For more technical details on the FHI-FEL see Ref. [19].

The VIS beam ($\lambda_{VIS} = 532$ nm) is generated by a frequency-doubled 1064 nm, amplified, table-top Nd:VAN laser (for details see Ref. [16]). It provides ~10 μs long macro pulses of ~4 mJ energy at 10 Hz, consisting of ~11 ps long micro pulses at a repetition rate of 55.5 MHz, i.e., with an identical pulse structure as the FEL. At the sample, the VIS beam has a diameter of about 600 μm. The temporal pulse overlap between VIS and FEL is based on an RF-based synchronization[20].

The generated SFG beam is detected using a 50× magnification long working distance objective (Mitutoyo M Plan Apo 50x), a 200 mm tube lens and a high-sensitivity, electron multiplied CCD camera (PI-MAX 4, Teledyne Princeton Instruments). The camera collects 10 FEL macro pulses per image. At each frequency, 10 images are recorded, leading to an acquisition time of 10 seconds per frequency. In order to separate the visible SFG light ($\omega_{SFG}$) from the initial green visible laser ($\omega_{VIS}$) a single bandpass filter (512nm X 30nm BP 93T 25D, Edmund optics) is placed in front of the objective. Two additional tunable bandpass filters (547/15, Semrock) are placed behind the objective. For more technical details on the microscope see Ref. [18].

In post-processing, at each frequency the 10 SFG images are averaged, background and cosmic-ray corrected. The images are then analyzed for spatial features, or integrated over all pixels to generate the SFG spectrum. The spectra are interpolated and divided by a reference spectrum to correct for edges of the transmission range of the tunable bandpass filters. All SFG spectra are measured at room temperature and under ambient conditions. Therefore, to account for air absorption of the IR beam, each measured SFG spectrum is additionally divided by an IR air absorption spectrum. An example of the post-processing procedure can be found in the SI (Section 1).

The *m*-plane AlN sample (Nitride Crystals, Inc.) is $0.4 \pm 0.03$ mm thick with a diameter of 15 mm. The surface orientation is (1-100) $\pm 0.5°$ and labeled in the unit cell in Figure **1b**. The crystal axes perpendicular to the optical axis are labeled $a_1$ and $a_2$, the optical axis is labeled *c*. SFG data was acquired for the *c* axis along *x* and *y* independently by rotating the sample by the azimuthal angle $\varphi$, see Figure **1a**.

### THEORETICAL MODEL

For comparison to the experimental data, SFG spectra are calculated analytically as a function of the IR frequency $\omega_{IR}$. As a second-order nonlinear process, the SFG signal is generated from the second-order nonlinear polarization[21,15]

$$\vec{P}^{(2)} = \overleftrightarrow{\chi}^{(2)} : (\hat{e}_{VIS} \overleftrightarrow{L}_{VIS})(\hat{e}_{IR} \overleftrightarrow{L}_{IR}), \quad (1)$$

where $\overleftrightarrow{L}_i$ is the tensorial Fresnel transmission coefficient, $\hat{e}_i$ is the unit polarization vector of the beam at $\omega_i$ with $i = \{IR, VIS\}$. Since the wavelength of the VIS beam stays constant during the experiment, the only frequency-dependent components of $\vec{P}^{(2)}$ are the Fresnel transmission coefficient of the IR $\overleftrightarrow{L}_{IR}$ and the second-order nonlinear susceptibility tensor $\overleftrightarrow{\chi}^{(2)}_{jkl}$.

The Fresnel coefficients $\overleftrightarrow{L}_{IR}$ are derived from linear optics with the individual simulation contributions shown Figure **2**. The IR dielectric function of AlN in the proximity of the optical phonon resonances is well-described by a Lorentz oscillator model:[22]

$$\varepsilon(\omega_{IR}) = \varepsilon_\infty \left( \frac{\omega_{LO}^2 - \omega_{IR}^2 - i\omega_{IR}\gamma_{LO}}{\omega_{TO}^2 - \omega_{IR}^2 - i\omega_{IR}\gamma_{TO}} \right). \quad (2)$$

Here, $\varepsilon_\infty$ is the high frequency dielectric constant. $\omega_{TO}$ and $\omega_{LO}$ denote center frequencies of the transversal and longitudinal optical phonon modes with their damping parameters $\gamma_{TO}$ and $\gamma_{LO}$, respectively. Using the ordinary and extraordinary phonon modes from Moore et al.[23], the dielectric function parallel ($\varepsilon_\parallel$) and perpendicular ($\varepsilon_\perp$) to the optical axis $c$ are obtained. Figure **2a** shows the respective calculated real (solid line) and imaginary (dotted line) parts.

The z-component of the ordinary (o) and extraordinary (e) complex wave vector $\vec{k}$ for an m-plane crystal is calculated by[24]

$$k_z^o(\omega_{IR}) = 2\pi\omega_{IR}\sqrt{\varepsilon_\perp(\omega_{IR}) - \sin^2\theta_i},$$

$$k_z^e(\omega_{IR}) = 2\pi\omega_{IR}\sqrt{\varepsilon_\parallel(\omega_{IR}) - \frac{\varepsilon_\parallel(\omega_{IR})}{\varepsilon_\perp(\omega_{IR})}\sin^2\theta_i}. \quad (3)$$

The real part of the wave vector represents the IR wave propagation constant, whilst the imaginary part determines its decay length. The spectral dependence of both quantities are plotted in Figure **2b**.

We measured two crystal orientations for the optical axis along the lab coordinates $x$ (azimuthal angle, $\varphi = 0°$) or $y$ ($\varphi = 90°$). Due to the in-plane anisotropy and angle of incidence of the FEL, experiments for $c \parallel x$ differ significantly from $c \parallel y$. For demonstration purposes, however, only the calculation for $c \parallel x$ is shown. The formulas and plots for $c \parallel y$ can be found in the SI (Section 2). For $c \parallel x$, the Fresnel transmission coefficients are calculated by[25]

$$L_{xx}(\omega_{IR}) = \frac{2k_z^e(\omega_{IR})}{\varepsilon_\parallel(\omega_{IR})k_z^i(\omega_{IR}) + \varepsilon_0 k_z^e(\omega_{IR})},$$

$$L_{yy}(\omega_{IR}) = \frac{2k_z^i(\omega_{IR})}{k_z^o(\omega_{IR}) + k_z^i(\omega_{IR})}, \quad (4)$$

$$L_{zz}(\omega_{IR}) = \frac{\varepsilon_\parallel}{\varepsilon_\perp}\frac{2k_z^i(\omega_{IR})}{\varepsilon_\parallel k_z^i(\omega_{IR}) + \varepsilon_0 k_z^e(\omega_{IR})},$$

with the incident wave vector $k_z^i(\omega_{IR}) = 2\pi\omega_{IR}\cos(\theta)/c_0$. $\varepsilon_0 \approx 1$ is the dielectric constant of air[26,27]. $c_0$ is the speed of light in vacuum. The imaginary and the real part of the Fresnel coefficients $L_{xx}$ and $L_{yy}$ are plotted in Figure **2c**. Since $L_{zz}$ is canceled out in the further calculation, it is not shown.

The second-order nonlinear susceptibility tensor $\overleftrightarrow{\chi}^{(2)}_{jkl}$ is a third-order tensor with the indices $j$ for the SFG, $k$ for the VIS and $l$ for the IR beam components in laboratory coordinates. The SFG polarization in propagation direction ($j = z$) does not contribute to the measured signal for homogeneous surfaces. For one exemplary geometry, where the two incoming beams are s-polarized, equation (1) becomes[15]

$$P^{(2)}_{SSS}(\omega_{SFG}) = \chi^{(2)}_{yyy} L_{yy}(\omega_{VIS}) L_{yy}(\omega_{IR}),$$

$$P^{(2)}_{PSS}(\omega_{SFG}) = \chi^{(2)}_{xyy} L_{yy}(\omega_{VIS}) L_{yy}(\omega_{IR}). \quad (5)$$

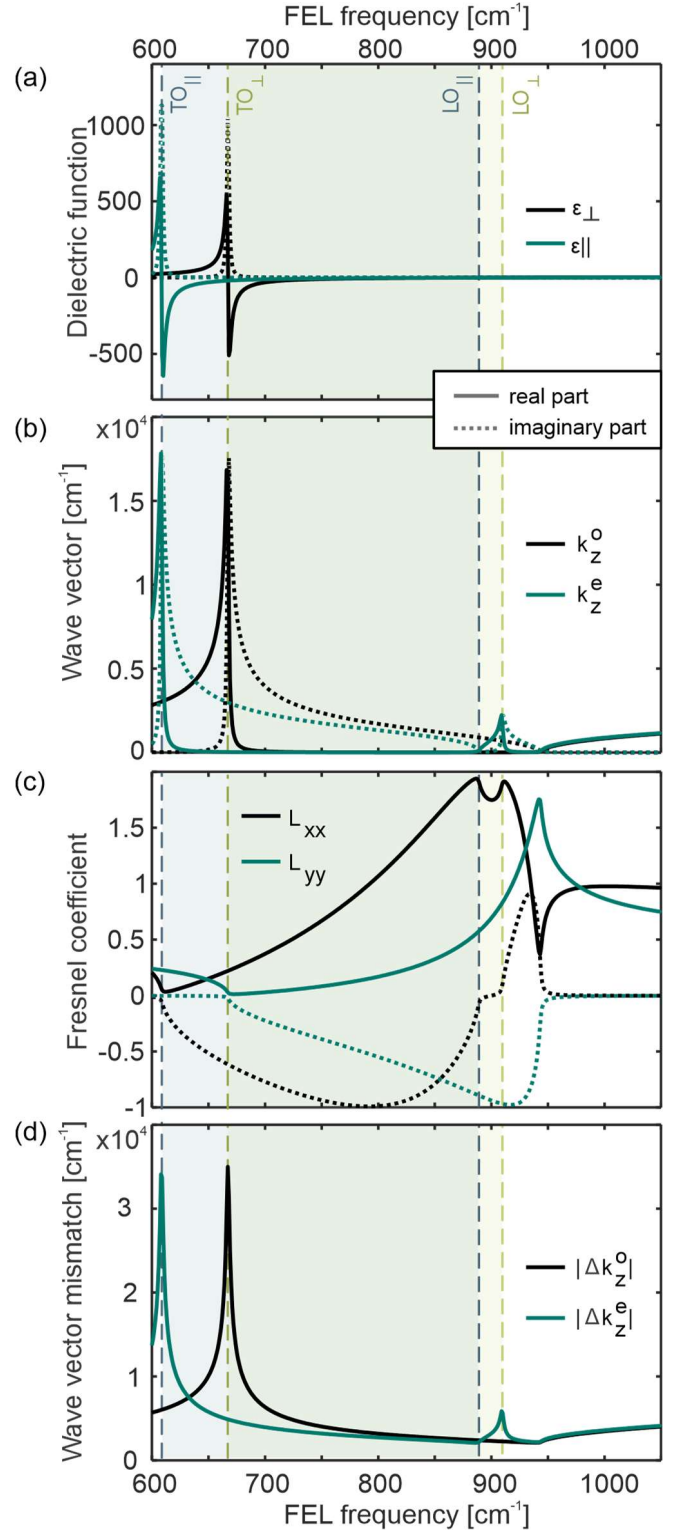

**Figure 2:** Individual steps of the theoretical simulation for $c \parallel x$: the real (solid) and imaginary (dashed) part of (a) the dielectric function, (b) the wave vector, and (c) the Fresnel factors. (d) shows the wave vector mismatch. The ordinary and extraordinary Reststrahlen bands are marked in light yellow and blue (overlap light green) shades, respectively.

Since the SFG signal is not measured polarization resolved, the final SFG intensity equals the sum $P^{(2)}_{SS}(\omega_{SFG}) = P^{(2)}_{SSS}(\omega_{SFG}) + P^{(2)}_{PSS}(\omega_{SFG})$. The SI (Section 3) show the resulting formulas for all 8 combinations.

As a second-order nonlinear process, broken inversion symmetry and therefore inheritably the crystal structure influence the SFG intensity. For every point group, specific $\bar{\bar{\chi}}^{(2)}$-elements vanish due to symmetry restrictions. For AlN with the point-group symmetry $C_{6v}$ (6mm), the non-vanishing $\bar{\bar{\chi}}^{(2)}$-elements read[21]:

$$\chi^{(2)}_{aca}, \chi^{(2)}_{aac}, \chi^{(2)}_{caa}, \text{ and } \chi^{(2)}_{ccc}. \quad (6)$$

Upon transformation from lab to crystal coordinates, the equations in (5) for all possible polarization combinations with $c \parallel x$ simplify to

$$P^{(2)}_{SS}(\omega_{SFG}) = \chi^{(2)}_{caa} L_{yy}(\omega_{VIS}) L_{yy}(\omega_{IR}),$$

$$P^{(2)}_{SP}(\omega_{SFG}) = \chi^{(2)}_{aac} L_{yy}(\omega_{VIS}) L_{xx}(\omega_{IR}),$$

$$P^{(2)}_{PP}(\omega_{SFG}) = \chi^{(2)}_{ccc} L_{xx}(\omega_{VIS}) L_{xx}(\omega_{IR}), \quad (7)$$

$$P^{(2)}_{PS}(\omega_{SFG}) = \chi^{(2)}_{aca} L_{xx}(\omega_{VIS}) L_{yy}(\omega_{IR}).$$

To include resonance contributions, each $\bar{\bar{\chi}}^{(2)}$-element is written as the sum of the non-resonant component $\chi^{(2)}_{NR}$ and a resonant component per phonon mode[15]:

$$\chi^{(2)}_{ijk} = \chi^{(2)}_{NR,ijk} + \frac{A^{(2)}_{ijk}}{\omega_{IR} - \omega_{TO} + i\Gamma_{TO}}, \quad (8)$$

where $A^{(2)}_{ijk}$ and $\chi^{(2)}_{NR,ijk}$ are used as fitting parameters.

The SFG intensity is obtained via:

$$I_{SFG}(\omega_{SFG}) \propto \left| \frac{(\hat{e}_{SFG} \bar{\bar{L}}_{SFG}) \vec{P}^{(2)}(\omega_{SFG})}{\Delta \vec{k}(\omega_{IR})} \right|^2. \quad (9)$$

Here, $\Delta \vec{k}(\omega_{IR})$ is the wavevector mismatch of the SFG. Due to in-plane momentum conservation ($\Delta k_x = \Delta k_y = 0$), the wave vector mismatch is reduced to its z-contributions. Contrary to previous SFG literature[21,15], in our measurement geometry the infrared and visible beam are counter propagating leading to the wave vector mismatch:

$$|\Delta k_z(\omega_{IR})| = |k_{SFG}(\omega_{IR}) + k_{VIS} - k_{z,IR}(\omega_{IR})|. \quad (10)$$

with $k_{VIS} = \frac{n_{VIS}}{\lambda_{VIS}}$ and $n_{SFG} \approx n_{VIS}$ being constant. Figure **2d** shows the wave vector mismatch.

**RESULTS AND DISCUSSION**

Figure **3a-d** presents the fitted simulation (solid lines) and the experimental spectra (dots) for four different configurations of VIS and IR polarization combinations, and two different sample orientations. The individual dots correspond to the integral over the entire field of view at a given frequency. Two exemplary images for VIS p IR s and $c \parallel x$ at 948 cm$^{-1}$ and 810 cm$^{-1}$ are shown in Figure **3e,f**.

In all measurements, a significant SFG signal is detected across the whole AlN Reststrahlen band spectral region. We distinguish depending on the polarization of the IR beam and the crystal orientation. There are three different scenarios with qualitatively different behavior: (i) Only ordinary features in the s-polarized geometry with $c \parallel x$ (black data in Figure **3c,d**) or in the p-polarized geometry with $c \parallel y$ (green data in Figure **3a,b**), (ii) Only extraordinary features in the s-polarized geometry with $c \parallel y$ (green data in Figure **3c,d**), (iii) Both ordinary and extra-ordinary features in the p-polarized geometry with $c \parallel x$ (black data in Figure **3a,b**). Note that the polarization of the VIS beam also influences the overall magnitude of the spectra, however, we focus on discussing the IR-frequency dependence first.

At high frequencies above 850 cm$^{-1}$, all spectra show distinct spectroscopic features, where the signal drops rapidly above the LO phonon frequency range. Notably for all cases (i)-(iii), the high-frequency edge of the Reststrahlen band is blue-shifted to the respective LO frequency by approx. 30 wavenumbers. This spectral shift emerges from the oblique incidence angle of the IR beam[26]. Individually, the exact position depends on the geometry: (i) is shifted relative to the ordinary LO phonon mode (Figure **3a,b** $c \parallel y$ and Figure **3c,d** $c \parallel x$), (ii) is shifted to the extraordinary LO phonon mode (Figure **3c,d** $c \parallel y$), (iii) shows features of both ordinary and extraordinary features, resembled in the shape of an extra dip in between the phonon modes (Figure **3a,b** $c \parallel x$). All these spectral features near the LO phonon modes are largely arising from linear optical effects, i.e., from the Fresnel coefficients and the wave vector mismatch, as exemplified for $c \parallel x$ in Figure **2**. The emergence of out-of-plane resonances for case (iii) in shape of additional ordinary features in our signal proves the importance of a tilted angle of incidence geometry for our method.

At low frequencies below 700 cm$^{-1}$, a small dip is observed at the TO phonon resonances, especially for geometries probing the ordinary TO phonon mode (Figure **3a,b** $c \parallel y$ and Figure **3c,d** $c \parallel x$). The exact spectral shape is determined by the competition of the linear-optical suppression near the TO phonons and the nonlinear enhancement arising from the $\chi^{(2)}$- resonances, as depicted in Figure **2**. Similarly, the shape of the spectra between the TO and LO resonances is determined by the interplay between linear and nonlinear contributions, in agreement to previous works on SHG[28]. The competition between the nonlinear and linear optical contributions strongly affects the fitting of the non-resonant $\chi^{(2)}_{NR,ijk}$ and the resonant $A^{(2)}_{ijk}$ parameters. Details on the fitting are given in the SI (Section 4). Note that especially at 750 cm$^{-1}$ some measured spectral features do not fit to the simulated data. We suspect that these divergences can result from residual air absorption.

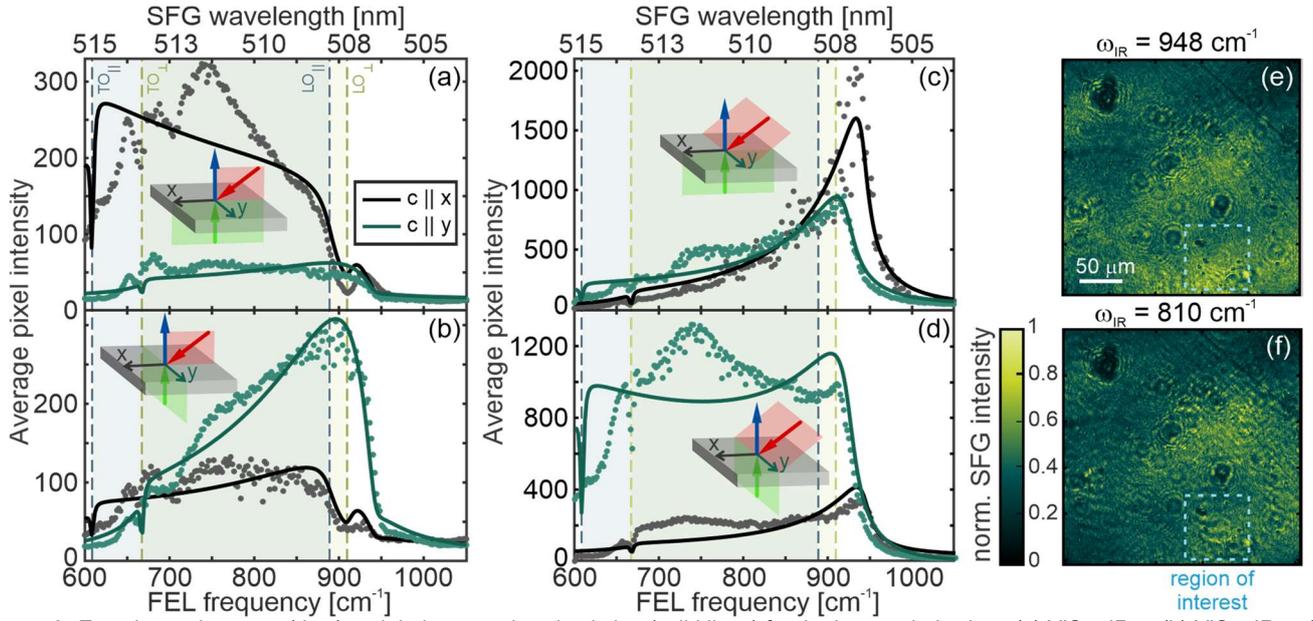

**Figure 3:** Experimental spectra (dots) and their respective simulation (solid lines) for the laser polarizations (a) VIS p IR p, (b) VIS s IR p, (c) VIS p IR s and (d) VIS s IR s, illustrated by the insets. The orientation of the optical axis is parallel to *x* (black) and *y* (green), respectively. The ordinary and extraordinary Reststrahlen bands are marked in light yellow and blue (overlap light green) shades, respectively. (e,f) SFG microscope images for the laser orientation VIS p IR s and the optical axis parallel to *x* at (e) 948 cm$^{-1}$ and (f) 810 cm$^{-1}$. The dashed box in (e) and (f) marks a region of interest, where polariton propagation is observed in (f), see text.

Globally, the different magnitudes of the $\chi_{NR}^{(2)}$-elements lead to laser polarization and sample orientation dependent signal intensities over the entire spectral range. Especially the polarization of the VIS beam leads to severe changes in the overall magnitude, while the spectral features may remain similar. For example, the pixel intensity in the $c \parallel x$ orientation in Figure **3c** is five-fold increased compared to Figure **3d**. In our current experiments, each configuration and sample orientation was measured individually, therefore the comparison of absolute intensities should be handled with care. However, in a potential multi-domain sample, the different sample areas and domains would be measured simultaneously. In principle, our spectro-microscopy approach enables spectroscopic analysis for different domains, down to individual pixels. In this scenario, an absolute comparison between the domain spectra will be possible. Furthermore, artifact pixels, e.g., from cosmic rays, can be excluded to improve the signal-to-noise-ratio. These benefits of spectro-microscopy make our method valuable for future domain imaging.

In order to demonstrate the microscopic capability of our method, Figure **3e-f** shows two exemplary SFG microscope images at two different IR frequencies. Linear microscopy images of the sample position are shown in the SI (Section **5**), which show a non-uniform surface structure due to dust particles and scratches on the sample. For all laser polarizations and sample orientations, here VIS p IR s, $c \parallel x$, fringes around the particles can be observed. However, at 948 cm$^{-1}$ (Figure **3e**) non-equidistant fringes result from linear optical interference (see Poisson's spot[29]), while at 810 cm$^{-1}$ (Figure **3f**, region of interest) hints of additional polaritonic waves with equidistant fringe spacing can be observed. The latter will be subject of an upcoming work. Here, we highlight the high spatial resolution of SFG spectro-microscopy, such that IR spectral information can be obtained at the diffraction limit of visible light.

In conclusion, we employed IR-VIS SFG spectro-microscopy in 8 unique geometries to show the interplay between linear and nonlinear optics impacting the spectral features. With this, the spectroscopic feasibility of our instrumentation for in-plane anisotropic aluminum nitride as a model system was proven, as we find the experimental spectra to agree well with the theoretical calculations. Additionally, our microscopic observations highlight the great advantage of spectro-microscopy, where morphology or local features of interest, e.g., polaritons or potentially domains, can be identified and converted to spectra by pixel-wise integration. In the future, this method will therefore provide spatially resolved spectroscopic SFG information in inhomogeneous systems such as domains and domain walls in ferroics.

## SUPLEMENTARY MATERIAL

The Supplementary Material contains more detailed information on the data analysis, the calculation for $c \parallel y$, the full theoretical calculation for both cases, the fitting procedure and the linear microscope pictures of the sample.

## ACKNOWLEDGEMENTS

We thank the whole FHI-FEL operator team, Sandy Gewinner, Marco de Pas and Wieland Schoellkopf for reliably providing us with lasing during our beamtimes. D. M. acknowledges funding by the Max Planck-Radboud University Center for Infrared Free Electron Laser Spectroscopy. S.F.M. acknowledges support by Deutsche Forschungsgemeinschaft (DFG, German Research Foundation, no. 469405347).

## CONFLICT OF INTEREST

The authors have no conflicts to disclose.

## DATA AVAILABILITY

The data that support the findings of this study are openly available in "Sum-Frequency Generation Spectro-Microscopy in the Reststrahlen Band of Wurtzite-type Aluminum Nitride - Experimental Data" at https://doi.org/10.5281/zenodo.11235361.